\documentclass[12pt]{article}
\usepackage[english]{babel}
\usepackage{a4wide}
\usepackage{latexsym}
\usepackage{amsfonts}
\usepackage{graphicx}
\date{}

\begin{document}

\title{\bf {\large{Symmetric Spaces of Exceptional Groups}}}
\author{\normalsize{Luis J. Boya \footnote{luisjo@unizar.es} \footnote{To be published
in the Proceedings of the  $27^{\rm {th}}$ International Conference on Group-Theoretical Methods in Physics. Yerevan,
 Armenia (2008).}} \\
\normalsize{Departamento de F\'{\i}sica Te\'{o}rica} \\
\normalsize{Universidad de Zaragoza} \\
\normalsize{E-50009 Zaragoza} \\
\normalsize{SPAIN}}

\maketitle

\begin{abstract}
We adress the problem of the reasons for the existence of $12$ symmetric spaces with
the exceptional Lie groups. The $1+2$ cases for $G_2$ and $F_4$ respectively are
easily explained from the octonionic nature of these groups. The 4+3+2 cases on the $E_{6,7,8}$ series
require the magic square of Freudenthal and, for the split case, an appeal to
the supergravity chain in $5, 4$ and $3$ spacetime dimensions.

\end{abstract}

\section{Introduction}

{\it {Symmetric Spaces}} were introduced and classified by E. Cartan in 1926.
They are riemannian spaces with an extra symmetry condition, namely
geodesic reflection being isometry; they turn out to be homogeneous spaces $S=G/K$.
This has as consequence that curvature is covariant
constant, $\nabla R=0$; that fact can be used as an alternative definition. Here we shall refer
mainly to the {\it {compact}} subclass; the standard source is the book by Helgason \cite{Hel}. From the
Lie algebra $L=L(G)$ point of view, there is an involutory automorphism with $L(K)$ as the fix point set.
Then if $P$ is a suplementary space, so $L(G)=L(K) \ \dot {+} \ P$, $[L(K),L(K)]=L(K), \ [L(K),P]=P, \ {\rm {and}} \ [P, P]=L(K)$.
A good recent reference for physicists in this context is \cite{Ugama}.
If $G$ and $K$ are compact, with Weyl\'{}s unitary trick $P \rightarrow iP$ we obtain the {\it {noncompact}} families of symmetric spaces.\\

    There are {\it {seven families}} of symmetric spaces
associated to the {\it {three division algebras}}, the reals $R$, the complex $C$ and the quaternions $H$,
as the transitive groups ($G$ above) in general are $O(n), U(n) $  and $ Sq(n)$,
and there are also {\it {twelve single}} cases.
The aim of this work is to understand the $12$ exceptions as extensions of the classical
families to the octonion algebra $O$. \\

     This study must be seen as a case of understanding exceptional cases in
mathematics: once one has classified objects, as e.g. in regular and exceptional classes, the question
arises to place exceptions in some context, to see why they are there. One can ask,
 for example, why there are exceptional Lie groups (like $E_8$), or sporadic (finite simple) groups (like the Monster group), or why some regular polytopes exist in four dimensions only (like the $24-$cell).
 Indeed, our problem has relation with other
  problems in Lie group theory: what are the {\it {real forms}} of (complex) Lie groups (classical
  or exceptional), which groups act {\it {transitively on spheres}}, and also with Berger\'{}s
  classification of the {\it {special holonomy}} manifolds; in this late case, for example, there are,
  besides the $3\cdot 2=6$ families, for the groups $SO, O; \ SU, U; \ Sq \ {\rm {and}} \ q(n):=[Sq(n)\times Sq(1)]/Z_2$,
 also two exceptions, again related to octonions, namely $G_2$ and Spin$(7)$. \\

  It is of course the lack of associativity of the octonion division algebra which makes the
  corresponding "series" finite in number. \\

   So we are adressing a purely mathematical question. We hope, however, that the issue
  would be relevant for physics: exceptional groups appear conspicuously e.g. in
   string or $M$-Theory \cite{Boy2}, and on general grounds one expects Nature to be described by a
   singular, non-generic mathematical structure, as Nature is unique (!). There are also recent
   applications of symmetric spaces to integrable systems and to random matrices. We shall make use
   of Supergravity in relation with symmetric spaces at the end of this report.\\

\section{Classification of Compact Symmetric Spaces}
The conventional (generic) symmetric spaces are very easy to describe. Let us write the series of groups
and inclusions for the Orthogonal ($O$), Unitary ($U$) and compact Symplectic ($Sq$) series:

\begin{equation}\label{eq:1}
O(n) \subset U(n) \subset Sq(n) \subset U(2n=m)\subset O(2m)
\end{equation} \

Roughly speaking, the four first classes of generic compact riemannian symmetric spaces are $U/O$, $Sq/U$,
$U(2)/Sq$ and $O(2)/U$.\\

NOTE: For us $Sq(n)$ is the {\it {compact}} form of the $C_n$ series of Cartan\'{}s list
of simple Lie groups; we shall use also (as above) the notation $q(n)$ for the ``nonunimodular" form
of $Sq(n)$. These should be understood as the ``unitary" groups with quaternion entries, or with
(double) unitary complex entries fixing an (extra) conjugation; ``q" is for quaternion, ``Sq"
for unimodular quaternion, of course.\\

    The first four uniparametric families of (compact) symmetric spaces are just the consecutive
quotients above:

\begin{equation}\label{eq:2}
SU(n)/SO(n), \ \   Sq(n)/U(n), \ \   SU(2n)/Sq(n),   \ \  {\rm {and}} \ \ SO(2n)/U(n).
\end{equation}
\begin{equation}
Dim \ \ (n-1)(n+2)/2,  \ \  n(n+1),   \ \  (2n+1)(n-1),  \  {\rm {and}} \   n(n-1).
\end{equation}
\begin{equation}
 Cartan  \ notation:  AI,  \ \   CI, \ \  AII, \ {\rm {and}} \ DIII.
\end{equation}

Besides, there are three biparametric families, with the same field,
 giving the grassmannians or collection of subspaces, namely
$K(p+q)/K(p)\times K(q)$, where $K:= R, C, H$ or groups $O, U$ and $Sq$:

\begin{equation}\label{eq:3}
SO(p+q)/SO(p)\times SO(q), \ \ SU(p+q)/SU(p) \times SU(q), \ \ Sq(p+q)/Sq(p) \times Sq(q).
\end{equation}

\noindent with dimensions $pq$, $2pq$ and $4pq$ respectively. In Cartan\'{}s notation they are
$BDI$, $AIII$ and $CII$.\\

These grasmannians are, as said, collection of subspaces; in particular, the important class of
projective spaces are reported here, as e.g.

\begin{equation}\label{eq:4}
RP^n = O(n+1)/O(n)\times O(1) = SO(n+1)/O(n) = S^n/S^0
\end{equation}

Similarly

\begin{equation}\label{eq:5}
CP^n = ... = SU(n+1)/U(n) = S^{2n-1}/S^1, \ \ HP^n= ... = Sq(n+1)/q(n) = S^{4n-1}/S^3
\end{equation}

The Lie groups themselves are also symmetric spaces, because one can write $ G  = [G\times G]/G_{diag}$.
But we do not consider them further. \\

Symmetric spaces have also a {\it  {rank}}, the dimension of the maximal totally
geodesic flat submanifold. It corresponds to the abelian (Cartan) subalgebra of the
Lie algebra of simple Lie groups. Symmetric spaces of rank one are, for example,
the projective spaces and the spheres, e.g. $S^n = SO(n+1)/SO(n)$ and $S^{2n-1}=SU(n)/SU(n-1)$,
as $SO(1)=SU(1)=1$. These spaces are also called {\it {two-point homogeneous spaces}}, Cfr. \cite{Hel}.
The free classical motion on them is superintegrable, maximally antiergodic (the orbits filling up
just a curve (or $1d$ manifold)), and quantization yields a discrete spectrum, depending on a
single quantum number. \\

There remain the twelve exceptional symmetric spaces, $S=G/H$, where $G$ is one of the five exceptional
groups. Cartan found the $1+2+4+3+2=12$ symmetric spaces associated respectively
 to the exceptional groups $G_2$, $F_4$, $E_6$, $E_7$ and $E_8$. To understand them, firts we have to
 see precisely the relation they have with the octonion division algebra.

\section{Generalities on Octonions and Related Groups}

    We start from the {\underline {reals}} $R$; it is an ordered field, $x<y$ makes sense, and there
are no automorphims, Aut$(R)$=1. Now in $R^2$ we define an $R-$algebra by $i^2=-1$, where $i=(0,1)$. Then
the complex numbers C are $z=x+iy$ (Gauss). Conjugation, norm and inverse are given as $\bar{z}=x-iy$, $N(z)
=\bar{z}z$, and for $ z \neq {0},\ z^{-1}=\frac {\bar{z}}{N(z)}$. Conjugation is the only continuous automorphism,
Aut$_c(C)=Z_2$; the field C is no longer ordered; both $R$ and $C$ are {\it {composition algebras}}, meaning the norm
verifies $N(z_1z_2)=N(z_1)N(z_2)$. We have the inclusions $R^2\supset S^1 \supset S^0$ for all complex numbers $C$, the unit complex numbers, and the imaginary units $\pm \ {i}$. \\

For the {\it {quaternions}} $H$ we proceed similarly (Hamilton, 1842). In $R^4$ define $i=(0,1,0,0)$ ,
 $j=(0,0,1,0)$ and $k=ij=(0,0,0,1)$ with $i^2=j^2=k^2=-1$, which requires $ij + ji =0$.
 Conjugation, norm and inverse are defined likewise as $q:=u + ix + jy + kz \equiv Re(q) + Im(q)$ and then
 $\bar{q}= Re(q) - Im(q)$, $N(q)=\bar{q}q$, and for $q \neq {0}, \ q^{-1}=\frac {\bar{q}}{N(q)}$. The inclusions now are $R^4 \supset S^3 \supset S^2$ for all quaternions $H$, the unit quaternions and the imaginary units. Conjugation now is {\it {anti}automorphism}, but arbitrary rotations of the imaginary units {\ {are}} the automorphims,
 Aut$(H)=SO(3)$. $H$ is an anticommutative (skew-)field. Note $N(q_1q_2)=\overline{q_1 q_2} \ q_1q_2=\bar {q_2} \bar {q_1}q_1q_2=N(q_1)N(q_2)$ \\

 For the {\it {octonions}} $O$ we proceed similarly (Greaves, 1842; Cayley, 1845). In $R^8$ define three
 independent antiinvolutory anticommutative units $i, j, k$ (e.g. with $i^2=j^2=k^2=-1$, $ij+ji=0$ etc). Now
 {\it {force}} the same properties for the products $ij, jk, ki; (ij)k$: this implies nonassociativity, namely
 $(ij)k=-i(jk)$, the most remarkable (for unusual) property of octonions.
 There are $1+$ ${3}\choose{1}$ + ${3}\choose{2}$ + $1 =8$ units, seven imaginary. Conjugation, norm and inverse are defined likewise; the automorphim group should be a subgroup of $SO(7)$: the inclusions now are $R^8\supset S^7
 \supset S^6$. So Aut$(O)$ is $G_2$, the first (smallest) exceptional simple Lie group, of dimension $14$ and rank $2$. The octonions form a composition division noncommutative nonassociative algebra; in particular $N(o_1o_2)=N(o_1)N(o_2)$ still holds. The concrete expression $(ij)k=-i(jk)$ is called {\it {alternativity}}, and does characterize the octonions: in general $[o_1o_2o_3]:=(o_1o_2)o_3
  - o_1(o_2o_3)$ is fully antisymmetric. For a further study, consult \cite{Baez}.\\

    For the reals $R$ we have the units $S^0=(\pm 1)=Z_2=O(1)$.
    For the complex $C$ the units are $S^1=\rm {exp}(i\phi )=SO(2)=U(1)$.
    For the quaternions $H$ we have $S^3= \rm {Spin}(3)=SU(2)=Sq(1)$.
    But for the octonions $O$ the sphere $S^7$ is {\it {not}} a group, only a parallelizable
    manifold (i.e. the tangent bundle is trivial). However, as shown by Ramond \cite{Ram}, the seven sphere
    becomes a group after stabilization by $G_2= \rm {Aut} (O)$, and becomes the group Spin$(7)$!

    \begin{equation}\label{eq:6}
                \rm {Spin}(7) \sim G_2 \vee S^7
    \end{equation}

    We need now some results from Lie group theory. Firts, any compact simple Lie group
    can be understood (Hopf, ca. 1940) as finitely twisted product of odd-dimensional spheres, symbolically

    \begin{equation}\label{eq:7}
      G \sim \Pi_j \odot S^{2j+1}
    \end{equation}

    \noindent where $j$ are called {\it {the exponents}}; for all this see \cite{Boy1}. For example. $SU(2)=S^3$ is known,
    $\rm {Spin}(4)=S^3\times S^3$ is true, but also $SU(3) \sim S^3 \odot S^5$, etc. Other result is that the exponents are palindromic (Kostant), i.e., the differences left-right are the same. For example, the exponents $j$ for Spin$(10)$ are $(1,3,4,5,7)$. \\

    Now we compare $Sq(1)=S^3$ and $q(1)=S^3\times_2S^3$ with $G_2 \sim S^3 \times S^{11}$ and Spin$(7) \sim S^3 \times S^7 \times S^{11}$. We shall call

    \begin{equation}\label{eq:8}
    S^7 \sim `Oct(1)',  \  \ G_2 \sim SOct(1), \ \ \rm{Spin}(7) \sim Oct(1)
    \end{equation}

    \noindent on the understanding that $`Oct(1)'$ is NOT a group. To justify the notation, we remark:
    $Sq(1) \rightarrow q(1) \rightarrow S^3$ corresponds to $SOct(1) \rightarrow Oct(1) \rightarrow S^7$
    which is true: $Spin(7)/G_2=S^7$. There are more examples, e.g. $Sq(1)^2= S^3 \times S^3$ corresponds to
    $Oct(1)^2=S^3 \odot S^7 \odot S^7 \odot S^{11}= \rm {Spin}(8)$. Now as $q(2) \sim Sq(1) \times Sq(2) \sim
    S^3 \odot S^3 \odot S^7$, it leads to $Oct(2) \sim S^3 \odot S^7 \odot S^{11} \odot S^{15}= \rm {Spin}(9)$! \\

    We can go up to $n=3$, but there is a surprise! $`Oct(3)'$ would have exponents $j: (1, 3, 5, 7, 11)$ , as corresponds to spheres (7, 15, 23) stabilized by (3, 11). But this combination is NOT palindromic! However, it becomes so by losing the $S^7$ sphere ((1,5,7,11) is OK!). Losing the 7-sphere is the unimodular ("S") restriction, and indeed we have concocted $F_4$:

    \begin{equation}\label{eq:9}
    F_4 \sim S^3 \odot S^{11} \odot S^{15} \odot S^{23} = SOct(3)
    \end{equation}

    \noindent which is the second exceptional Lie group, with dimension $52 = 3+11+15+23$ and rank 4 (number of spheres)! \\

    We stop here for the moment. As summary, we have ``shown'' the following "equivalences''
    $`Oct(1)'=S^7, \ G_2 = {\rm {Aut}}(O)\sim S^3 \odot S^{11}= SOct(1), \ Oct(1)=S^3 \odot S^7 \odot S^{11} = {\rm {Spin}}(7), \ Oct(1)^2 = {\rm {Spin}}(8), \ Oct(2)= {\rm {Spin}}(9), \ SOct(3)=F_4 $.

\section{Symmetric Spaces of Exceptional Groups: $G_{2}$ and $F_{4}$ }

    We are ready now no address the exceptional symmetric spaces. As we say that
$G_2=SOct(1)$ and Spin$(7)=Oct(1)$, we generalize the $Sq(1)$ case: its only
symmetric space is of $q/U$ type, namely $Sq(1)/U(1) \approx S^3/S^1=S^2=CP^1$.
Hence the octonionic equivalent must be

\begin{equation}\label{eq:10}
 SOct(1)/q(1) \approx G_2/SO(4) \approx \frac {S^3 \odot S^{11}}{S^3 \odot S^3} \approx HP^2
 \end{equation} \\

 It is the {\it {only}} symmetric space with $G_2$. We do not care about the ``Octonionic'' forms for
 Spin$(7, 8, 9)$, so our next case will be $F_4$. Recalling $F_4=SOct(3)$, there are now two
 possibilities: $Oct/Sq$ and $Oct(3)/Oct(1)\times Oct(2)$; both are present: from

 \begin{equation}\label{eq:12}
 O(2) \subset U(2) \subset q(2) \subset Oct(2) = {\rm {Spin}}(9)
 \end{equation}

 \begin{equation}
 O(3) \subset U(3) \subset q(3) \subset SOct(3) = F_4
 \end{equation} \

 \noindent we get $SOct(3)/Oct(2) = F_4/{\rm {Spin}}(9) = OP^2$, i.e., the Moufang (octonionic, projective) plane.
 But also $SOct(3)/q(3) = F_4/Sq(3) \times Sq(1)$, the other symmetric $F_4$-space. \\

 There is another case, namely $Oct(2)/Oct(1)^2 \approx Spin(9)/Spin(8) = S^8 = OP^1$, but
 this is already accounted for, e.g. as $S^8=SO(9)/SO(8)$.\\

 Are there more cases? One proves that all other putative cases have been already accounted for.
 For example, from $U(2)/U(1)=S^3$ and $q(2)/q(1)=S^7$, one would expect $Oct(2)/Oct(1)=S^{15}$. Indeed
 ${\rm {Spin}}(9)/{\rm {Spin}}(7)=SO(9)/SO(7)\equiv SO(9)/SO(7) \times SO(1)$, {\it {dej\`{a} vu}}, as it is
 an orthogonal grassmannian (and $S^{15}$ plays the role of a Stiefel manifold); it corresponds to the principal
 bundle ${\rm {Spin}}(7) \rightarrow {\rm {Spin}}(9) \rightarrow S^{15}$, an exceptional transitive action
 on spheres: in fact, Spin$(9)$ is the largest Spin group acting {\it {trans}} in any maximal sphere! \\

 So we have $S=G_2/SO(4)$ with dim $S=8$; it is Cartan\'{}s  $G$ space. And $S'=F_4/{\rm {Spin}}(9)$ with
 dim $16$; it is Cartan\'{}s $FII$ space. Finally $S''=F_4/Sq(3)\times Sq(1)$ with dim $28$; it is
  Cartan\'{}s $FI$ space. \\

 \section{Magic Square and the $E$-Series: the Other Symmetric Spaces}
 For the $E$-series we anticipate our construction of the symmetric spaces
 is not so clear-cut. One of the reasons is, undoubtedly, that our
 understanding of these groups is still incomplete, inspite the efforts of
 Freudenthal, Tits, and others. We shall resort to two contructions for these
 groups: (1), The {\it {Magic Square}} of Freudenthal, Rosenthal and Tits (1956),
 and (2) an appeal to physics, in the existence of the chain of maximal
 supergravities, from spacetime dimensions $11$ to $5, 4, 3$. \\

 Recall the inclusions already seen $O(n) \subset U(n) \subset q(n) \subset U(2n) \subset O(4n)$, which
 holds for any $n$; it might be ``filled in'' for $n=2$ and $n=3$ as follows:

 \begin{equation}\label{eq:13}
 O(2) \ U(2)  \ Sq(2) \ Oct(2)= {\rm {Spin}}(9)
 \end{equation}
 \begin{equation}
 U(2) \ U(2)^2 \ U(4) \ Oct_C(2) = {\rm {Spin}}(10)
 \end{equation}
 \begin{equation}
 Sq(2) \ U(4) \ O(8) \ Oct_H(2) = {\rm {Spin}}(12)
 \end{equation}
 \begin{equation}
 {\rm {Spin}}(9) \ {\rm {Spin}}(10) \ {\rm {Spin}}(12) \ {\rm {Spin}}(16)=Oct_O(2) \\
 \end{equation} \\

 Also, the primordial Magic Square is

 \begin{equation}\label{eq:13}
 O(3) \ U(3)  \ Sq(3) \ Oct(2)= F_4
 \end{equation}
 \begin{equation}
 U(3) \ U(3)^2 \ U(6) \ E_6
 \end{equation}
 \begin{equation}
 Sq(3) \ U(6) \ O(12) \ E_7
 \end{equation}
 \begin{equation}
F_4 \ \ E_6 \ \ E_7 \ \ E_8 \\
 \end{equation} \\

 The last row/column must be understood as $SOct(3)=F_4$, extended by the
 complex ($E_6$), the quaternions ($E_7$) and the octonions ($E_8$).
 Of course, the right way to understand these groups is as automorphism
 groups for the $3\times 3$ hermitian Jordan algebras over the bi-rings
 octonions$\otimes$reals $=$ octonions, octonions$\otimes$complex,
 octonions$\otimes$quaternions and octonions$\otimes$octonions. See e.g. \cite{Freuden}\\

 Now for $E_6$: we have the three inclusions
 \begin{equation}\label{eq:15}
 F_4 \subset E_6, \ \ U(6) \subset E_6, \ \ O(10) \subset E_6
 \end{equation}
 which generates  {\it {three}} of the 4 cases for $E_6$. For $E_7$ we have also
 the three inclusions
 \begin{equation}
 E_6 \subset E_7, \ \ O(12) \subset E_7 \ ({\rm {repeated!}})
 \end{equation}
 which generates {\it {two}} of the 3 cases for $E_7$. For $E_8$ we have also
 two more cases:
 \begin{equation}
 E_7 \subset E_8, \ \ O(16) \subset E_8 \ ({\rm {repeated!}})
 \end{equation}
 which generates the {\it {two}} cases for $E_8$. \\

 The folowing eqs. summarizes the 3+2+2 cases:

 \begin{equation}\label{eq:16}
 E_6/F_4, \ E_6/(SU(6)\times SU(2)), \ E_6/SO(10)\times U(1) \; E_7/E_6\times U(1), \ E_7/O(12) \times Sq(1);
 \end{equation}
 \begin{equation}
 E_8/E_7 \times Sq(1), \ E_8/O(16).
 \end{equation}
  \begin{equation}
 {\rm {with \ dimensions}} \ 26, \ 40, \ 32; \ \ 54, \ 64; \ \ 112, \ 128
 \end{equation}
 corresponding to Cartan\'{}s $E IV$, $EII$, $EIII$; $EVII$, $EVI$; $EIX$ and $EVIII$. \\

 Before looking for the remaining two cases (one for each $E_6, \ E_7$) we pause to
 remark that the three following cases appear in physics:

 \begin{equation}\label{eq:17}
 F_4/{\rm {Spin}}(9)=OP^2 \ \ ({\rm {M \ Theory}}), \ \ E_6/{\rm {Spin}}(10)\times SO(2) \ \ ({\rm  {F \ Theory}}) \ \ E_8/SO(16) \  ({\rm {Superstrings}})
 \end{equation}\\

 The astute reader would have notices that the two remaining cases are the {\it {split}} form of the $E_{6,7}$ groups, namely the quotients by $Sq(4)$ and $SU(8)$ bzw., which correspond to the split form divided by the maximal compact
  subgroup of the mother group. Where are they? First, recall the notations

 \begin{equation}\label{eq:18}
 E_6^{compact}=E_{6(-78)}, \ E_6^{split}:=E_{6(+6)} \approx R^{42} \times Sq(4) \ ({\rm {dim}} \ 36)
 \end{equation}

We have no better way to understand the remaining symmetric spaces that to draw the sequences of groups and
 subgroups appearing in Supergravity, from dim $11$ down to dim $3$, in such a way that the scalars form a sigma-like
 model. In fact, these pairs groups-subgroups are the {\it {split}} form:

\begin{equation}\label{eq:19}
E_8 \supset E_7 \supset E_6 \supset D_5 \supset A_4 \supset ...
\end{equation}
\begin{equation}
O(16) \supset SU(8) \supset Sq(4) \supset Sq(2)^2 \supset Sq(2) \supset...
\end{equation}

\noindent which corresponds, in the descent from the $11$-dim Supergravity of \cite{Scherk} to

\begin{equation}\label{eq:20}
3d,\ \ 128 \ scalars. \ \ 4d, \ \ 35+35 \ scalars. \ \ 5d, \ \ 42 \ scalars. \ \ 6d, \ \ 25 \ scalars, \ \ 7d, \ 14 \ scalars...
\end{equation}

It happens that the subgroup sequence corresponds exactly to the maximal compact subgroup
of the main group sequence (I do not know why this is so).\\

This gives us {\it the remaining two } symmetric spaces: for $E_6$ we have: $Sq(4)$ with the traceless four-form
 makes up $E_6$, with ($36+42=78$). For $E_7$ we have: $SU(8)$ with the four-form makes up $E_7$, as $63+70=133$. Finally, although we already obtained that, for $E_8$ we have: $SO(16)$ with the Spin irrep make up $E_8$, with $120 + 128 = 248$. In this connection see \cite{Adams}. The first two were the missing exceptional symmetric spaces, corresponding to Cartan\'{}s $EI$, for $E_6/Sq(4)$,
 and $EV$ for $E_7/SU(8)$. \\

 Notice the sequences of groups $G(n), \ G(n)^2, \ G(2n)$ and $O(n), \ U(n), \ Sq(n) $ appear repeated
 many times. To quote a single example, the chain of first Spin groups is $Spin(1)=O(1), \ Spin(2)=U(1), \
 Spin(3) = Sq(1), \ Spin(4)=Sq(1)^2=q(1), \ Spin(5) = Sq(2), Spin(6) = SU(4)$, etc. It reminds one of Bott\'{}s
  periodicity.

 \section{Conclusions}1) The {\it twelve} exceptional symmetric spaces are
 obtained in a natural way, from pairs group/subgroup, both related to the octonions! \\
 2) For $G_2$ and $F_4$, the arguments are unequivocal and unavoidable. \\
 3) For the $E$ series we present two arguments: one is the natural generalization of the
 inclusions $O\subset U \subset Sq \subset Oct$, but the second argument makes use of a series of
 $E$- groups and subgroups for which the evidence is physical, not mathematical! This undoubtedly
 reflects the fact, already alluded to, that our understanding of the $E$ series is still incomplete.\\
 4) During the writeup of our talk, the paper by Ferrara appeared \cite{Ferrara}, which exploit the relation
 among {\it irreducible Riemannian globally symmetric spaces} and {\it supergravity theories} in $3$, $4$ and $5$
 space-time dimensions, and therefore there is some overlap with our considerations on the last two
 exceptional symmetric spaces. \\

 {\bf {Acknowledgements.}} I thank the organizers of the $27^{\rm {th}}$
 ICGTMP meeting at Yerevan (Armenia), especially G. Pogosyan, for the very stimulating
Conference. I thank the participants  A. Solomon, F. Toppan and the belgian group (U. of Ghent) for discussions.

 \vfill \eject

\end{document}